\documentclass{aa}
\usepackage{graphicx}
\usepackage{epstopdf}
\usepackage{txfonts}
\usepackage{xcolor}
\usepackage{natbib}

\bibpunct{(}{)}{;}{a}{}{,}

\DeclareMathOperator{\sech}{sech}
\usepackage{newunicodechar}
\newunicodechar{°}{\ensuremath{^\circ}}
\usepackage[export]{adjustbox}

\def\apgt{\ {\raise-.5ex\hbox{$\buildrel>\over\sim$}}\ }
\def\aplt{\ {\raise-.5ex\hbox{$\buildrel<\over\sim$}}\ }
\def\am{AM~CVn}   
\def\ams{AM~CVns}
\def\sna{SN~Ia}

\def\mwd{$M_{\rm WD}$}

\newcommand{\mdot}{$\dot{M}$}
\newcommand{\ms}{$M_\odot$}
\newcommand{\msun}{$M_\odot$}

\newcommand{\porb}{\mbox {$P_{\rm orb}$}}
\newcommand{\pmin}{\mbox {$P_{min}$}}
\newcommand{\mch}{$\mathrm{M_{Ch}}$}
\newcommand{\myr}{\mbox {~${\rm M_{\odot}~yr^{-1}}$}}
\newcommand{\ace}{\mbox {$\alpha_{ce}$}}
\newcommand{\al}{\mbox {$\alpha_{ce}\times\lambda$}}
\newcommand{\pyr}{\mbox {{\rm yr$^{-1}$}}}
\def\pc3{\mbox {{\rm pc$^{-3}$}}}

\begin{document}

\title{He-star donor AM CVn stars and their progenitors as LISA sources}

   \author{W.-M. Liu
          \inst{1}
          \and
          L. Yungelson \inst{2}
          \and
          A. Kuranov \inst{3}
          }
   \institute{\inst{1}School of Physics and Electrical Information,
   Shangqiu Normal University,  Shangqiu 476000, China,
   \email{liuwmph@163.com}\\
              \inst{2}Institute of Astronomy, Russian Academy of
              Sciences,
              48 Pyatnitskaya str., Moscow 109017, Russia\\
              \inst{3}Sternberg Astronomical Institute, Moscow State University,
               14 Universitetsky pr.,  Moscow 119992, Russia
               }

   \date{Received  (month) (day) (2022); accepted (month) (day) (2022) }


  \abstract
 { Ultracompact cataclysmic variables (CVs) of the AM CVn type are deemed to be  important verification sources for the future space gravitational  wave detectors such as the Laser Interferometer Space Antenna (LISA).}
   {We model the present-day Galactic population of AM CVn stars with He-star donors. Such a population has long expected to exist, though only a couple of candidates are known.   }
   {We applied the hybrid method of binary population synthesis (BPS) which combines a simulation of the population of immediate precursors of \am\ stars by a fast BPS code with subsequent tracking of their evolution by a full evolutionary code.   }
   {The model predicts that the present birthrate of He-donor AM~CVn stars in the Galaxy is $4.6\times 10^{-4}$~\pyr\  and  the Galaxy may harbour  $\simeq$112\,000  objects of this class  which have orbital periods $P \aplt 42-43$~min.
  The foreground confusion limit and instrumental noise of LISA prevent the discovery of longer periods systems  in gravitational
  waves.  We find that  about 500 He-star AM~CVns  may be detected by LISA with signal-to-noise ratio
  (S/N)>5  during a 4 yr mission.
  Within 1~Kpc
 from the Sun, there may exist up to 130  He-star AM~CVns with the periods in the same range, which may serve as verification binaries, if detected in the electromagnetic spectrum.
In the Milky Way, there are also $\simeq$ 14\,800 immediate precursors of \am\ stars.
 They are detached systems with a stripped low-mass He-star and a white dwarf companion, out of which
about 75 may potentially be observed by LISA during its mission.
}
   {}

   \keywords{Gravitational waves -- binaries: close -- cataclysmic variables -- evolution -- subdwarfs -- white dwarfs}

 \maketitle

\section{Introduction}
\label{sec:intro}
AM CVn stars are a small group of
ultracompact cataclysmic binaries (CVs) with He-rich (He white dwarf or stripped He star) donors and carbon-oxygen (CO) white dwarf (WD) accretors. Currently, about 70 definite and candidate \am\ stars are known, see
\citet[][Table 1]{2018A&A...620A.141R},
\citet{2016MNRAS.462L.106W,2020ApJ...905...32B,2021PASJ...73.1375K}, and \citet{2021AJ....162..113V}.
Estimated orbital periods of AM~CVns range from 5.4~min. to 67.8~min. Their
evolution is driven by gravitational waves radiation
\citep[GWR,][]{1967AcA....17..287P}. Detailed reviews of them were published by
\citet{2010PASP..122.1133S} and \citet{2018A&A...620A.141R}.

 \citet{1987SvA....31..228L} estimated that binary WDs may be the strongest GWR
sources  detected using lasers in space.  \citet{1996ConPh..37..457H} recognised that  \am\ stars are also expected to be detectable
by space GWR antennas. Along with massive black hole binaries and extreme or intermediate mass ratio inspirals, compact binaries
are among the main objects expected to be observed by the planned space GWR observatories such as the Laser Interferometer Space Antenna
\citep[LISA, see][and references therein]{2022arXiv220306016A},
Taiji, and TianQin \citep{gong2021concepts}.

The special importance of \am\ stars for GWR  astrophysics stems from the fact that some of them belong to the
so-called verification binaries (or guaranteed sources) for
space GWR detectors
(S. Phinney 2001, unpublished), \citep{2006CQGra..23S.809S,
2009CQGra..26i4030N,2018MNRAS.480..302K}, and \citet{2020PhRvD.102f3021H}.
It is expected that they will be discovered
rather soon  after the beginning of the missions because of a strong
signal due to the proximity to the Sun; additionally, thanks
to the knowledge of mass ratios of components, orbital periods and
distances  from observations in the electromagnetic spectrum, they will
serve for testing and calibrating detectors.

Theoretical modelling and observations suggest three formation channels for
 \am\ stars. A detailed analysis of them
may be found, for instance, in \citet{nele05} and \citet{2014LRR....17....3P}. Here, we recall only basic details of these channels.

The double-degenerate \citep[DD,][]{1967AcA....17..287P} channel envisions formation of a binary harbouring
a CO  WD and a less massive He WD companion. If the system is sufficiently tight,
the He WD may overflow its Roche lobe in a time shorter than the Hubble time due to angular  momentum loss
(AML) via GWR, and under certain conditions stable mass exchange is expected to commence
\citep[see also][]{1979AcA....29..665T,nele01b,2004MNRAS.350..113M}.
In the DD channel, the binary may evolve from $\porb \approx$ (2-3) min.
at the RLOF to $\porb \sim $ 1~hr.

In the single degenerate   channel (SD)
\citep{1972ApJ...175L..79F} and \citep{1986A&A...155...51S}, a stripped `semidegenerate' He-star may accompany CO WD.
In this case, RLOF by the progenitor of the donor should occur before He exhaustion in its core
\citep{1987ApJ...313..727I}. At the contact, \porb\ is
several dozen minutes, depending on the masses of components and the extent of exhaustion of He in the core of the future donor.
The binary first  evolves to \pmin $\sim$10~min., which is attributable to the
change in sign of the $M-R$ relation when the thermal timescale of the donor becomes substantially longer than the timescale
of angular momentum loss by GWR \citep[see][for more details]{2008AstL...34..620Y}.
The scenario of formation of He-donor  \am\ stars was extensively discussed, for example, by  \citet{1996MNRAS.280.1035T,nele01b,2006LRR.....9....6P,2010PASP..122.1133S,
2014LRR....17....3P,2020ApJ...904...56G}, and \citet{2021ApJ...922..245B}.

Following \citet{nele01b}, the objects that formed via an SD channel are classified as AM CVns after passing \pmin. However, as noted by \citet{2021ApJ...922..245B}, already before reaching \pmin, relatively high-mass subdwarf donors may transfer matter from (0.2 -- 0.3)\,\msun\ outer He shells that contain the ashes of CNO burning. In this case,  accretion disks spectra resemble the spectra typical for AM CVns.

In the SD channel, when the donor mass decreases to
(0.02 -- 0.03)\ms, it begins to cool, its thermal timescale becomes shorter than the  mass-loss timescale, it becomes more degenerate, and
the $M-R$ relation gradually merges with that for the cool WD
\citep[see][for a detailed discussion]{2021ApJ...923..125W}.
The formation of a particular \am\ system via a DD or SD channel may be  inferred from the abundances of elements and their
ratios, especially N/He, N/C, N/O, O/He, and O/C  \citep{2010MNRAS.401.1347N}; however, the derivation of abundances is a formidable task.

The evolved donor channel \citep{tfey85} is, in fact, the standard scenario of formation of CVs,
in which the donor overflows the Roche lobe when the hydrogen abundance in the core becomes $\aplt 0.1$. The donor becomes a star with an almost H-depleted core and a thin H envelope, which later may be lost.
Hypothetical \am\ stars, which formed via this  channel, typically have
$\porb\apgt$30~min. and very rarely may evolve to \porb$\sim$10 min.
\citep{2003MNRAS.340.1214P,2015ApJ...809...80G,2016ApJ...833...83K,2018sas..conf..157Y,2021ApJ...910...22L}.
Strictly speaking, this channel does not produce `classic' \am\ stars, since an abundance of H at the surface of the donors hardly decreases below $10^{-4} - 10^{-3}$, while spectral lines of H should be observed for an abundance exceeding
$10^{-5}$\ \citep{2009A&A...499..773N}. The parameters of the well-studied system Gaia14aae are apparently matched  best by the evolved donors channel, but no H
is observed in its spectrum \citep{2018MNRAS.476.1663G}.

For completeness, we note that about a dozen so-called HeCVs with an enhanced He/H abundance ratio have  \porb\ below  conventional
$P_{\rm orb, min}$
\citep{2012MNRAS.425.2548B,2015gacv.workE..25B,2015ApJ...815..131K,2022ApJ...925L..22L}.
Some of them may finish their evolution in the sub-minimum periods' range or bounce (change the sign of $\dot{P}$) and return to  \porb$\apgt$70-80 min, retaining H
in the envelopes. Envelopes of other HeCVs may contract after the mass of them drops
below a certain minimum. Then  they experience
a detached stage of evolution and join  the family of DD \ams, due to continuing angular momentum losses. The rest of the H envelopes may be expected
to be lost very soon after RLOF, when \mdot\ is high.

Observational data suggest that the population of \am\ stars is totally dominated by the objects that formed via the DD channel. Currently, only several
candidate \am\ stars with a He-star donor are  known -- SDSS J0926+3624 \citep{2011MNRAS.410.1113C},
ZTFJ1637+49, and ZTFJ0220+21 \citep{2021AJ....162..113V}.

The first binary population synthesis (BPS) studies of AM CVn stars, including both DD and SD channels, were performed by
\citet{1996MNRAS.280.1035T,nele01b}, and \citet{nele04}. Later studies, as a rule, were aimed at DD.
This is mainly related to the apparent scarcity of observed He-donor \ams\ and the badly understood consequences of accretion of He onto WDs at the expected accretion rates,
especially for rotating WDs.  Even so, the studies of He-star \ams\ may be important; this is because if they exist, but they are simply not recognised due to selection effects, they may increase
the sample of  LISA verification binaries.

For the present paper, we modelled the Galactic population of the stripped He stars with WD companions which later form
AM CVn stars with He-star donors. We modelled the formation of them and followed their evolution to RLOF and through
the mass-transfer stage to the instant when the mass of accretors reached \mch\  or the mass of the donor
decreased to (0.02 -- 0.03)\,\ms,
the limit imposed by the evolutionary code used. We estimate the number of \am\ stars and their direct precursors and evaluate the number of \am\ stars that may be detected in GW with a signal-to-noise ratio (S/N) > 5 by space-detector LISA during a 4 yr long mission. We present our model in Section 2. In Section 3 the results of the modelling are presented. We discuss the results and summarise our conclusions in Section 4.

\section{The model}
\label{sec:model}
\subsection{Population synthesis}
For the modelling, we applied  the hybrid
BPS method \citep{2012JPhCS.341a2008N,2014MNRAS.445.1912C,2015ApJ...809...80G}:
the evolution of binaries  up to the formation of precursors of AM CVn systems, WDs accompanied by He stars,
was computed by means of an updated
fast analytic BPS code BSE \citep{htp02}\footnote{After the update, our version of BSE did in fact  become rather similar to the latest published version of this code \citep{2020A&A...639A..41B}).}, while their further evolution was simulated using a grid of pre-computed full evolutionary tracks.
The advantage of hybrid population synthesis over other algorithms of BPS,
implemented, in particular, in BSE, is that the latter are
usually based on analytic approximations to
the evolutionary tracks for relatively well-explored single stars.  Description of the evolution of close binaries, for which systematic studies are much scarcer, since they are more complicated, is often based on
`educated guesses'. This is particularly true
for the second RLOF in the systems with compact accretors. Furthermore, as test runs show, BSE does not reproduce the evolution of semi-detached systems with  He donors, which gradually become more degenerate.

The crucial assumptions were as follows.
The initial mass function (IMF) of primary components followed the Salpeter law ($dN/dM\sim M^{-2.35}$) in the mass range
$1 \le M_1/M_\odot \le 100$.
A flat distribution of
mass ratios of components $q=M_2/M_1\le 1$  in the [0.1,1] range \citep{1989Ap.....30..323K} was assumed
\footnote{\citet{kor17} found that the flat distribution provides better agreement of the model with the number of WDs observed in the immediate vicinity of the Sun than
often used $f(q) \propto q^{-1}$.}.
The stellar binarity rate was set to  50\%,
that is to say two-thirds
of all stars are binary components.
The initial distribution of close binaries over orbital periods
was accepted after
Sana et al. (2012): $f(\log \porb) \propto \log \porb^{-0.55}$.
For common envelopes (CE) treatment, we applied the energy-balance formalism of
\citet{web84} and \citet{dek90} with  `CE efficiency'
parameter \ace=1 and binding energy parameter $\lambda$\ values from \citet{2011ApJ...743...49L}.

Evolutionary tracks for tracing further evolution were
computed with the updated P.P. Eggleton evolutionary code STARS
\citep[][year 2006 version, \it{private communication}]{1971MNRAS.151..351E}; for more details, readers can refer to  \citet{2008AstL...34..620Y}. All computations were carried out for metallicity Z=0.02.
The STARS code lacks  opacity tables that would allow one to compute models with masses $\aplt$0.02\,\ms\ corresponding to \porb$\apgt$(42--43 min.)\footnote{This is the problem common to many evolutionary codes \citep[see, e.g. discussion in][]{2021ApJ...923..125W}.}.  By  coincidence, at \porb\ which slightly exceeds  40 min.,  it is impossible to detect a GWR signal of \am\ stars by LISA, due to the foreground and antenna noise (see Fig.~\ref{fig:f_hc}  below).

\subsection{Computation of gravitational waves' strain}
Via BPS,  we found the masses of WD ($M_{\rm WD}$),
stripped He stars ($M_{\rm He}$), and orbital
periods just after the end of RLOF or common-envelope phases
$P_i$ for binaries,  in which a He star may later initiate
stable mass transfer.
For the mission lifetime of LISA $T$ = 4 yr, the characteristic strain of an inspiraling binary can be calculated  as \citep{1987thyg.book..330T}
\begin{equation}
h_{\rm c}\approx 3.75\times 10^{-19}\left(\frac{f_{\rm gw}}{1~\rm mHz}\right)^{7/6}\left(\frac{\mathcal{M}}{1~M_{\odot}}\right)^{5/3}\left(\frac{1~\rm Kpc}{d}\right),
\label{eq:hc}
\end{equation}
where $f_{\rm gw}[Hz]=2/\porb$[s] is the GW frequency, $d$ is the distance to the object,
and
\begin{equation}
\mathcal{M}=\frac{(M_{\rm WD}M_{\rm He})^{3/5}}{(M_{\rm WD}+M_{\rm He})^{1/5}}
\label{eq:chirp}
\end{equation}
is the so-called chirp mass.

Helium-donor \am\ stars are young objects ($t\lesssim 2$\,Gyr) and belong predominantly to the thin disk population  \citep{2018A&A...620A.141R}. Therefore, to  determine $d$, we assumed that the space distribution of progenitors of AM~CVn stars in the Galaxy may be described as
\begin{equation}
\rho \propto \exp{\left(-\frac{R} {R_d}\right)}\sech^2{\left(\frac{z}{z_d}\right)},
\label{eq:space}
\end{equation}
where $1 \le R \le 16$ Kpc is the galactocentric radial distance, $R_d=2.5$\,Kpc is the characteristic radial scale,
$z$ is the distance to the Galactic plane, and
$z_d$=0.3\,Kpc is the characteristic scale height of the disk \citep{2008ApJ...673..864J}.
We did not consider the inner region of the Galaxy with $R\leq$1\,Kpc, hosting a `bulge/bar', where young stars are absent
\citep{2020MNRAS.499.2340R,2020MNRAS.499.2357J,2022arXiv220507964J}. The volume of  the `excluded' region is quite conservative, taking the complicated structure of the inner region of the Milky  Way into
account \citep[e.g.][]{2016A&A...587L...6V}.
The Galactic thin disk age was set to 10\,Gyr.

The current number of proto- and \am\ systems was obtained by
convolving the birthrates of model systems with their lifetimes and star formation rate (SFR).
considered to be constant
over the past 2-3 Gyr and equal to 2\,\myr\
\citep{2011AJ....142..197C,2015ApJ...806...96L}.
The birthrate of AM CVns precursors for SFR 1\,\myr\ may be found as $\nu=C \times N_{AM}/N_{BSE}$, where $N_{AM}$ is the number of precursors obtained by evolving $N_{BSE}$ initial systems by BSE and C=0.045 is the percentage of systems with $M_1 \ge 1$\,\ms\ in the [0.1,100]\ms\ range (for the Salpeter IMF).
The number of stars that spend time $\Delta t_k$ in the rectangular cell of ($f,h_c$) space
limited by [$f_i, f_i+\Delta\,f, (h_c)_j,(h_c)_j+\Delta\,h_c$] (per initiated star), where $\Delta\,f$ and $\Delta\,h_c$ are the steps of the regular grid,
 respectively, was computed as follows:
\begin{equation}
\Delta N(t_k,t_k+\Delta t_k) =
\frac{\sum\limits_{l=1}^{N_{STARS_{i,j}}}\Delta t^{k,l}}{\Delta
t_k}\times \frac{1}{N_{BSE}},
\label{eq:conv}
\end{equation}
where $0 \le \Delta t^{k,l} \le \Delta t_k$ is the duration of the AM\,CVn stage for the system $l$,
$N_{STARS_{i,j}}$
is the number of systems in the
given cell according to the STARS
 grid, and
$N_{BSE}$ is the number of systems initialised in BSE.
We used $N_{BSE}=10^5$.

\section{Results of computations}
\label{sec:results}
\subsection{Formation of He-donor AM~CVn stars}

\begin{figure}[]
\vskip -0.35cm
\center
\includegraphics[width=1.0\columnwidth]{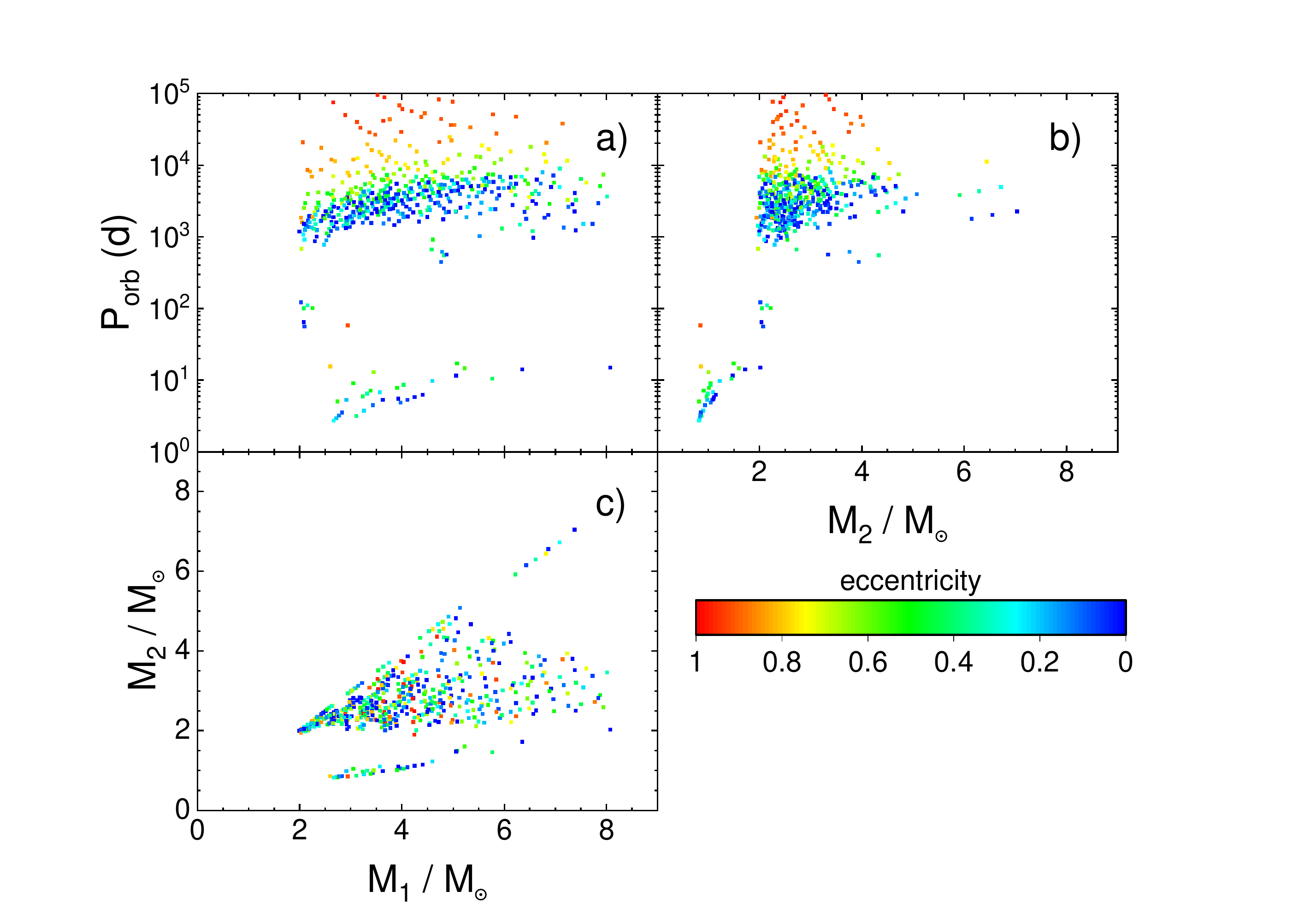}
\vskip -0.5cm
\includegraphics[width=1.0\columnwidth]{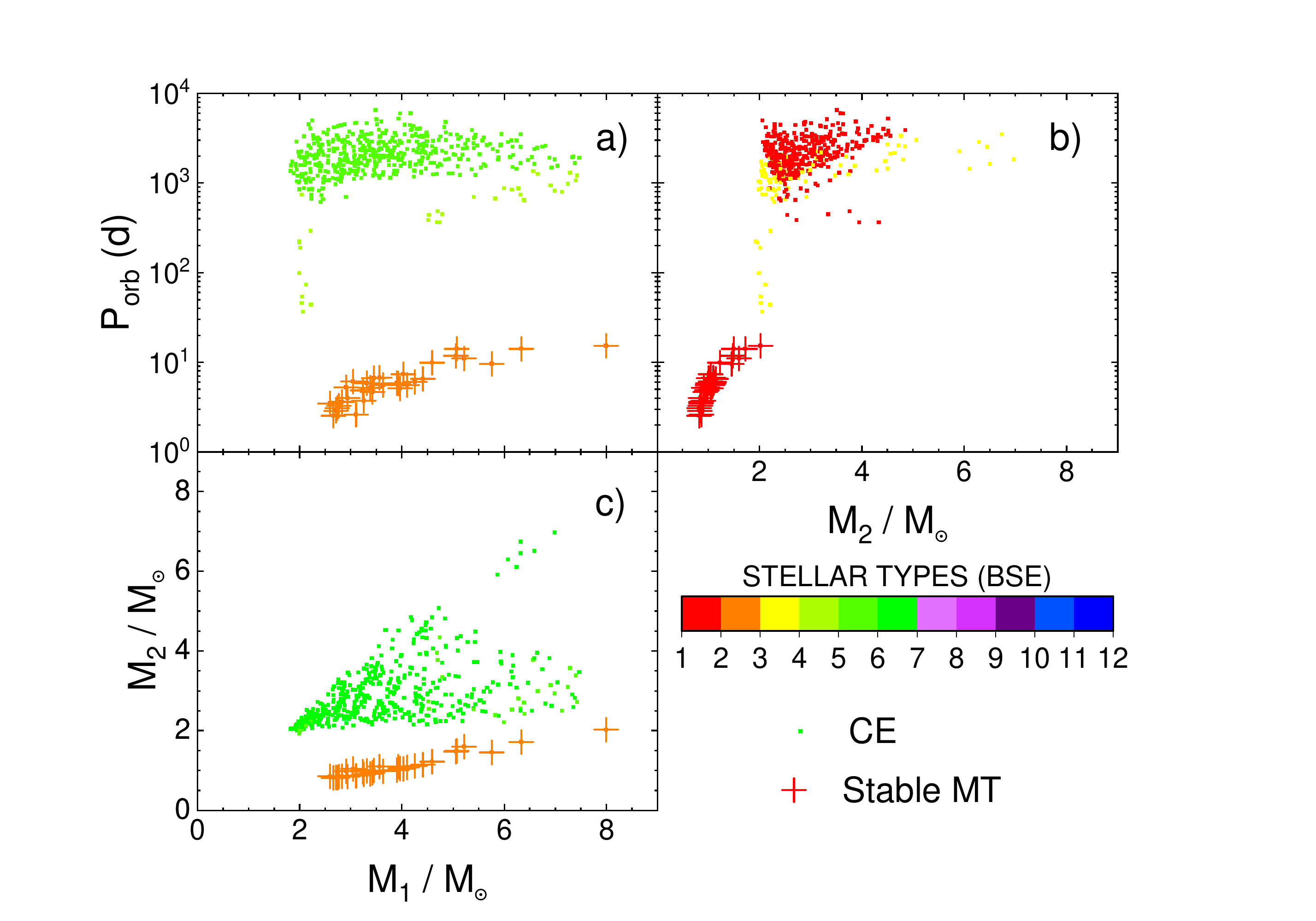}
\vskip -0.5cm
\includegraphics[width=1.0\columnwidth]{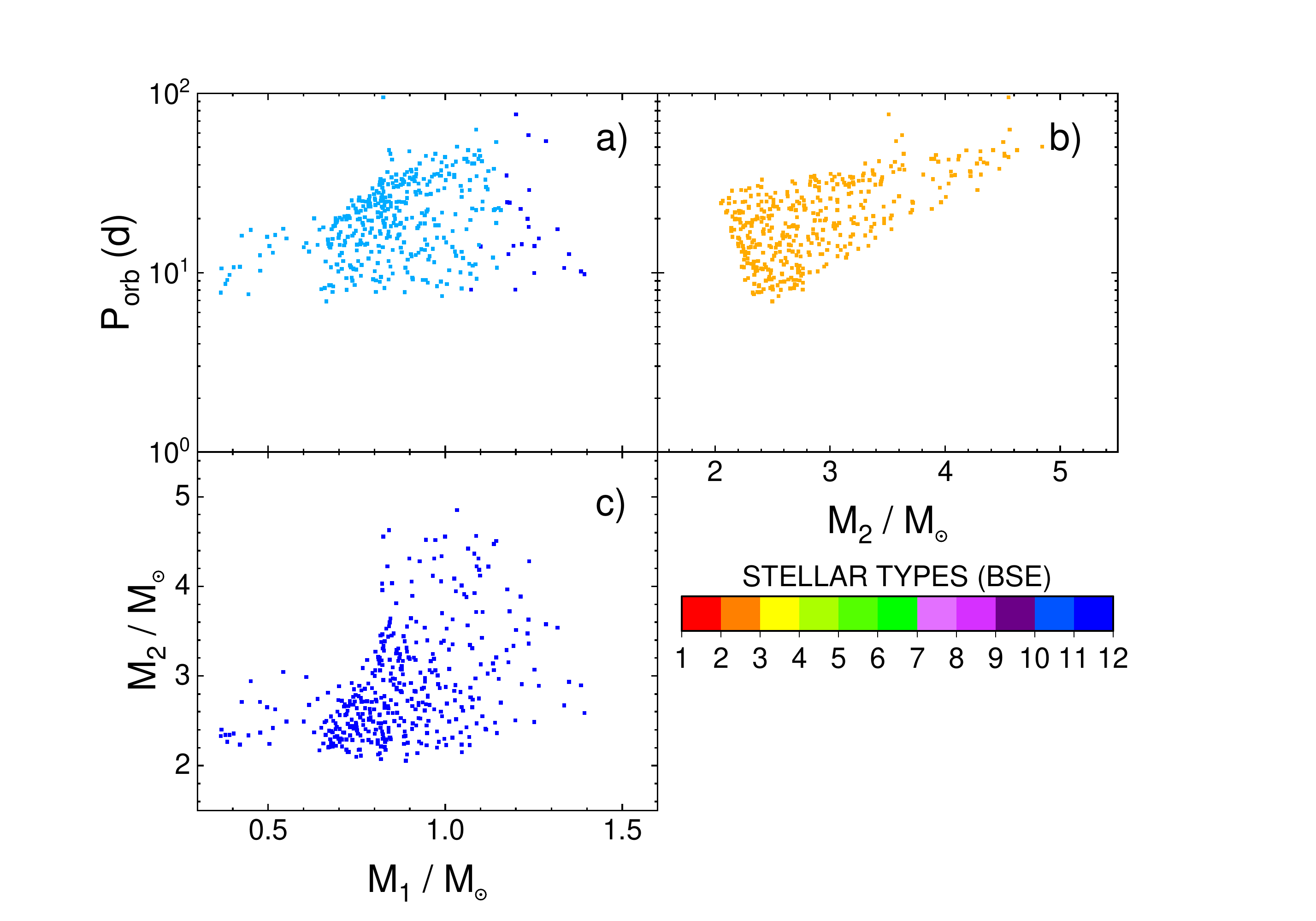}
\caption{Evolution of the relations between the parameters and types of components in the  progenitors of AM CVn stars. Upper panel --  precursors of AM CVns generated by BSE at ZAMS.
a) Relation between the masses of progenitors of WDs and initial \porb;
b) relation between the masses of progenitors of He stars and
\porb;
c) relation between the masses of progenitors of WDs and He stars.
Middle panel -- the binaries immediately prior to the first RLOF.
Subpanels a, b, and c are the same as in the upper panel.
Dots represent the binaries which pass through CE,
crosses indicate the systems with stable RLOF.
Lower panel -- the binaries prior to the second RLOF.
a) Relation between the masses of WDs ($M_{\rm WD}$) and \porb;
b) relation between the masses of pre-He stars and \porb.
c) relation between $M_{\rm WD}$ and the masses of pre-He stars.
The colour scale in the upper panel shows orbital eccentricities; in the middle and lower panels, colours  indicate the types of systems according to BSE: 1 -- MS star, 2 -- Hertzsprung gap star, 3 -- first-ascent red giant, 4 -- core He burning giant, 5 -- star at E-AGB, 6 -- TP-AGB star, 7 -- naked  He star, 8 -- He star in the Hertzsprung gap, 9 -- naked He-(sub)giant, 10 -- He  WD, 11 -- CO WD, 12 --ONe WD.
}
\label{fig:evol}
\end{figure}

Modelling the evolution  of $10^5$ stars in BSE resulted in the production of 524 progenitors of He-star AM CVns.
Corner plots in  Fig.~\ref{fig:evol} show the relations between the masses of components and orbital periods of initiated systems and  stellar types of binary components at ZAMS,  prior to the first RLOF in the system, which results
in the formation of WD components, and prior to the second RLOF, which results in the formation of stripped He components.
We note that some initiated systems have extremely large orbital eccentricities, but in the course of evolution
they circularise. While the first RLOF for some systems proceeds stably,
the second one almost always involves the formation of common envelopes.
\begin{figure}[ht!]
\vskip -0.8 cm
\hskip -1.3cm
\includegraphics[scale=0.484]{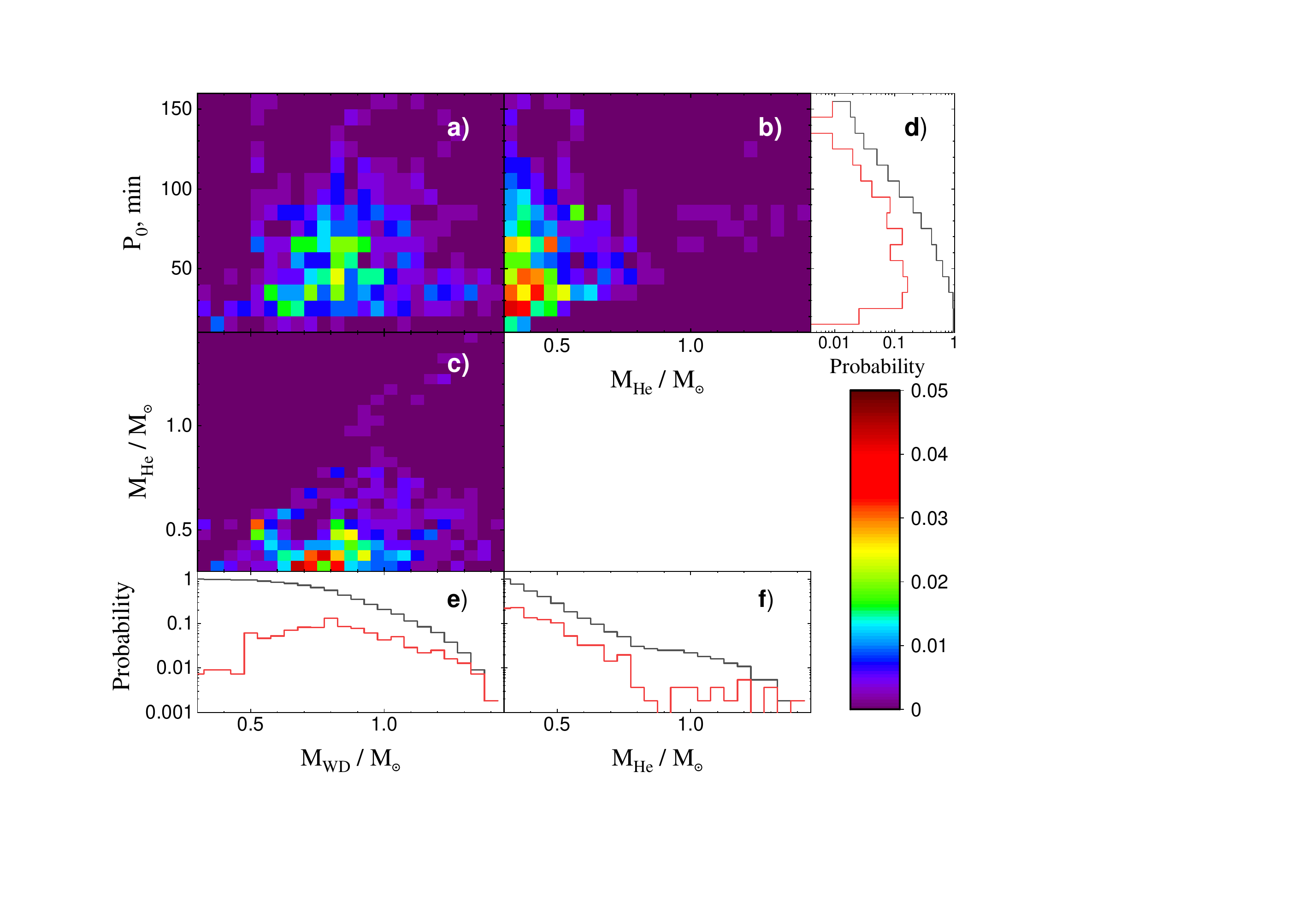}
\vskip -1.3cm
\caption{Relations between the parameters of detached He-star+WD systems which will evolve into  He-star AM~CVns.
The subpanels show the relations between the masses of WDs
and post-CE \porb\ (a),
between the masses of the nascent stripped He star and \porb\ (b),
between the masses of WDs and stripped star components (c), the
differential (red line) and cumulative (black line) distributions over \porb\ (d),
the differential and cumulative distributions over $M_{\rm WD}$ (e),
and the differential and cumulative distributions over masses of He stars (f).
The plots in the panels are normalised to unity.}
\label{fig:mwd_mhe_p}
\end{figure}

Distributions of the parameters of the pairs of naked He-stars+WD  after the second RLOF
is shown as a grid in Fig.~\ref{fig:mwd_mhe_p}.
The BPS results suggest that He stars are predominantly the lowest mass ones, from $M_{min}\approx$0.32\,\ms\ to $\approx$0.4\,\ms, but in some very rare cases they may be as massive as about 1.2\,\ms. Accretor masses are predominantly in the 0.65\,\ms\ to 1\,\ms\,range. The presence of relatively massive donors implies that even in the case of not very efficient accretion, WDs  may accumulate \mch, as suggested by \citet{2005ASPC..334..387S}.
The distribution is dominated by the binaries that formed with
\porb$\aplt$100\,min., but sometimes \porb\ attain 150\,min.  In wider systems, He in the cores of stars may be
exhausted before RLOF and they end their lives as WDs and merge with companions due to
AML via GWR.
In low-mass donors ($M_{\rm sdB,0} \aplt 0.4 -0.5$\ms) and more massive donors which overflow Roche lobes when He abundance in the cores is still $Y_c \apgt 0.3$,  He burning is quenched
almost immediately after RLOF and they transform into \am\ stars.
In more massive donors ($M_{\rm He,0} \apgt 0.55$\,\ms), if  $Y_c \aplt 0.3$ at RLOF, core He burning continues. When He in the centre is almost exhausted, the stars contract and  burn remainders of He.
Expansion after exhaustion of He results in a merger with
their companions \citep[for details, see the case of the (0.65+0.8)\,\ms, $P_{\rm orb,0}$=90\,min. system in][]{2008AstL...34..620Y}.

Figure~\ref{fig:mwd_mhe_p} provides information on the space of the initial parameters of immediate progenitors of He-star AM CVn systems and their birthrate.
We estimate the current Galactic birthrate of He-donor \am\ stars as $\approx 4.6 \times 10^{-4}$~\pyr.  This number is commensurate with the value $2.7\times 10^{-4}$\,\pyr\ found by \citet{nele04} for
the case of high-mass of the layer of accreted He, necessary for its
detonation and prevention of formation of  an AM~CVn star
 (see further discussion below).

\subsection{Population of He-donor AM CVn stars detectable by LISA }
\begin{figure}
\includegraphics[width=0.688\columnwidth,angle=-90]{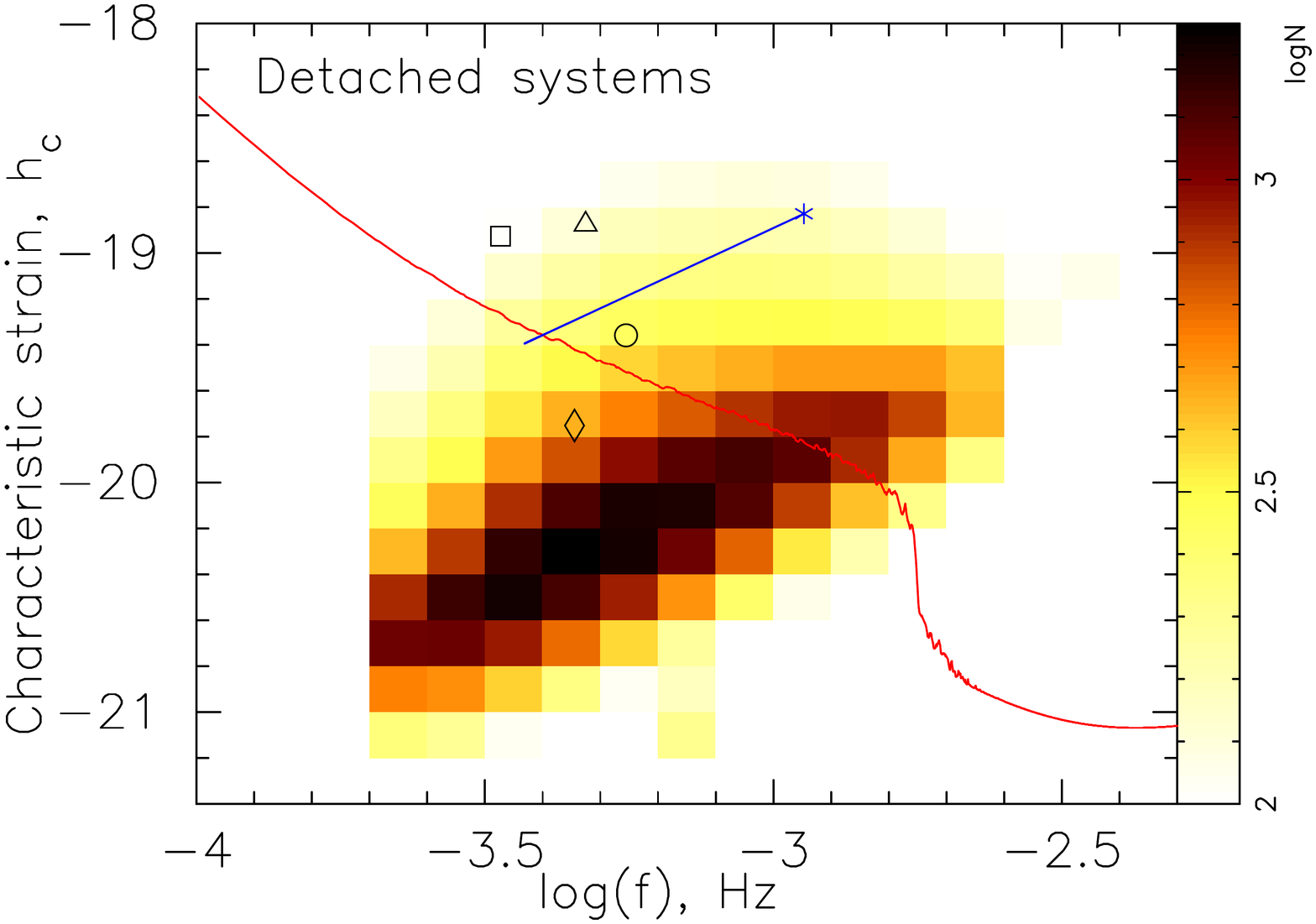}
\vskip -0.1 cm
\includegraphics[width=0.688\columnwidth,angle=-90]{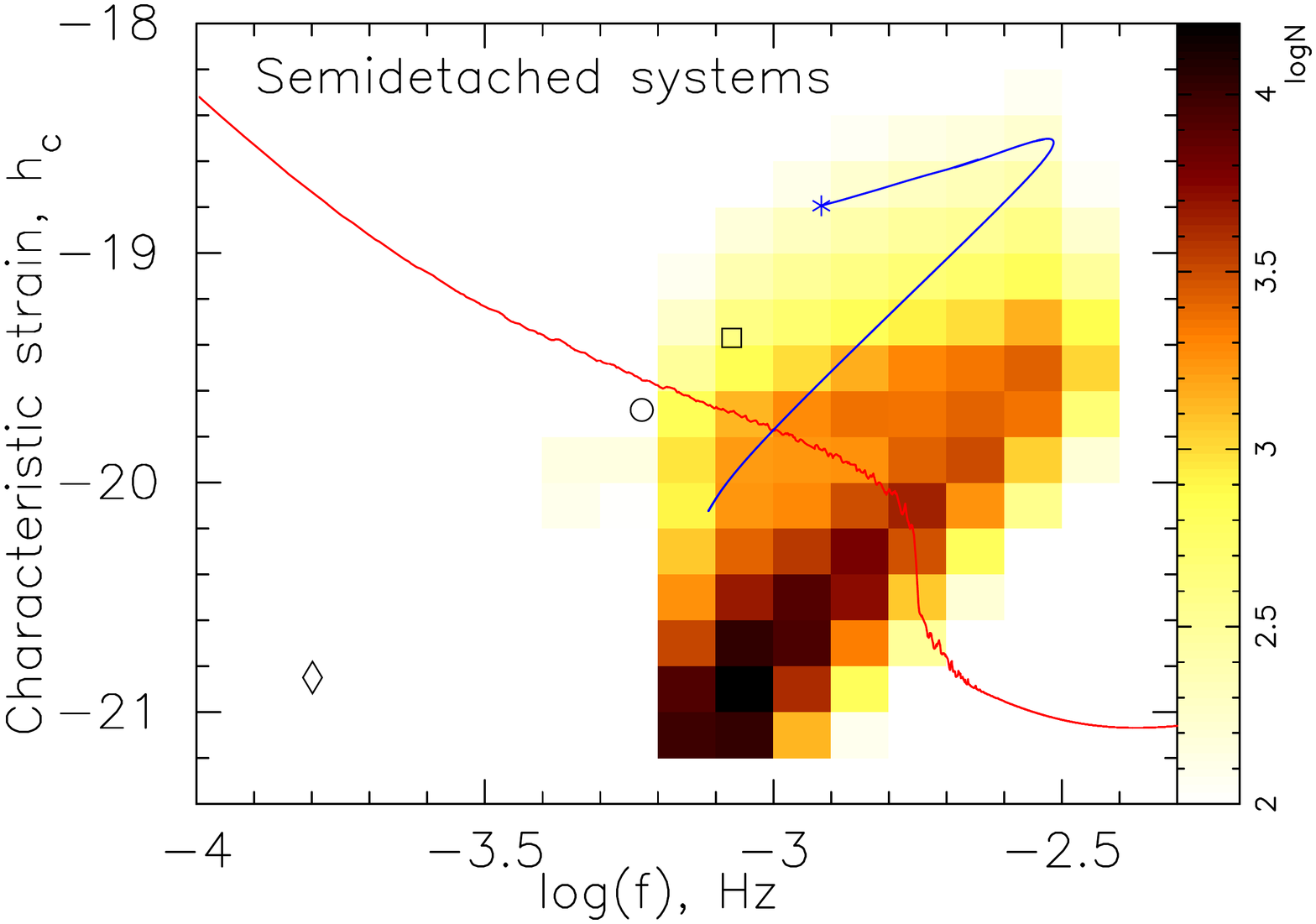}
\caption{Upper panel -- detached systems in the frequency - characteristic strain plane. Open circle, square, triangle, and diamond show positions of detached sdB+WD binaries
PTF~J2238+7430,
HD265435,
CD-30$^{\circ}$11223, and
OW~J0815–3421
respectively.
Solid red line shows
confusion limit due to binary WD  and LISA instrumental noise \citep{2022MNRAS.511.5936K}. Blue line shows
the pre-contact evolutionary track for the system with initial parameters
${\rm M_{He}=0.43}$\,\ms, ${\rm M_{WD}=0.87}$\,\ms, \porb=90\,min., placed at $d=1$\,Kpc. Position of the system at RLOF is marked by an asterisk. Lower panel -- semi-detached systems in the frequency - characteristic strain plane. Open circle, square, and diamond show positions of semi-detached sdB+WD systems ZTF~J2130+4420,
ZTF~J2055+4651, J1920-2001.
Blue line  -- continuation of the track shown in the upper panel in semi-detached stage.
See the text in \S~4.3 for objects references.}
\label{fig:f_hc}
\end{figure}
Based on the information provided in  Fig.~\ref{fig:mwd_mhe_p}, we computed a grid of 275 tracks with different combinations of
$M_{\rm WD,0}$, $M_{\rm He,0}$, and $P_{\rm orb,0}$ with the STARS code.
For computation of the $f-h_c$ relation, every computed system was assigned 20\,000 random  positions in the thin disk of the Galaxy. The $h_c$ obtained were then taken into account with the weight 1/20000.
Some cells  shown in Fig.~\ref{fig:mwd_mhe_p} contain up to five systems, and thus, altogether, we
had 365 tracks. In the case of several systems in a  cell, the computation of $h_c$  accounting for random
positions was repeated appropriately.

Figure \ref{fig:f_hc} shows the present-day distributions of detached pre-\am\  and \am\ systems
in the $f-h_c$ plane. It is mainly defined by the distances to the objects and lifetimes
of stars in the given frequency range. Evidently, the darkest shades are `populated' by
distant and long-living stars with low-mass donors.
The possibility of detecting  signals from most  systems is limited by the sum of the `confusion limit'
that is formed by the signals of the population of unresolved detached close DWD and AM CVn stars
\citep{eis87,1997CQGra..14.1439B,2000ApJ...537..334H,2001A&A...365..491N}
and LISA instrumental noise, which prevents the  detection of gravitational signals from binary WDs, unless they are very strong. We applied the
`observations-driven' confusion limit
\citep{2021PhRvD.104d3019K,2022MNRAS.511.5936K}\footnote{The estimate was carried out for LISA arms of 2.5\,Gm.}, based on the results of the studies of the local DWD population using large samples of objects.
The GW foreground dominates at $f\aplt$1.8\,MHz, while LISA sensitivity (for S/N >5)
defines detectability at higher frequencies (see Fig.~\ref{fig:tracks}).
the foreground is dominated by detached WDs and AM CVn stars do not play a noticeable role in its formation
\citep{2000ApJ...537..334H,2001A&A...365..491N}.
This justifies the  use of the confusion limit inferred from the observations of detached WDs only.

As it is mentioned above, our evolutionary code does not allow one to
compute the evolution of stars less massive than (0.02 -- 0.03)\,\ms\
(\porb$\apgt$(42-43)\,min.). We find that the Galaxy harbours about 112\,000 He-donor \am\ stars with  \porb$\aplt$(42-43)\,min.  However,
as it is clearly seen in Fig.~\ref{fig:f_hc},
for most systems the periods exceeding 42-43 min. are longer than the periods at the confusion limit.

It is clear from Fig. \ref{fig:f_hc} that, currently, most \am\ stars should have
masses of donors below 0.02\,\ms\ and relatively massive accretors ($\apgt 0.7$\,\ms, unless evolution is strongly non-conservative).

In order to illustrate our conclusions, we plotted in Fig.~\ref{fig:f_hc}
an evolutionary track for the system
with initial parameters ${\rm M_{He,0}=0.43}$\,\ms, ${\rm M_{WD,0}=0.87}$\,\ms, and ${\rm P_{orb,0}}$=90\,min., assuming that
its distance to the Sun is 1~Kpc. Such a system was chosen since its track  both starts and ends below the
foreground+sensitivity line.
We split the track into `detached' and `semi-detached' parts.
The system remained detached for about $\Delta t\approx$36~Myr. The
He star overflows its Roche lobe
when \porb$\approx$20.2 min.  Before  this, the binary
might be observed as an sdB+WD system. Several such possible proto-AM~CVn systems with well-measured parameters are known and indicated in Fig.~\ref{fig:f_hc}. The minimum \porb\ of the binary is 11.46 min., and it is
reached at  $t\approx42$~Myr.

Our exemplary binary becomes undetectable in GW when its orbital frequency declines to 1~mHz (${\rm P_{orb}} \approx  33.3$~min.) at $t\approx131$~Myr. At this
$t$, the mass  exchange rate is
$\approx1.4\times10^{-9}$~\myr, meaning that the system should still be bright in optical.
Later, \mdot\ rapidly declines. The code breaks at $t\approx496$\,Myr\ after the formation of a He-star + WD system,
when ${\rm P_{orb}} \approx 43.2$~min. and ${\rm M_{He}}\approx0.023$\,\msun.

\section{Discussion and conclusions}
\label{sec:disc}
We have presented the  results of the first study of He-star \ams\  and their immediate precursors by  the hybrid BPS.  The advantage of the method is more precise tracking of semi-detached binaries than by analytic approximations. However, the physics
implemented in the evolutionary code (opacity tables) restricts the range of orbital frequencies of AM CVns available for the study.

\subsection{Comparison to other studies}
We estimated that the birthrate of the Galactic thin disk He-donor \am\ stars is
4.6$\times10^{-4}$\,\pyr, and the number of objects with
\porb$\aplt(42-43)$\,min. is $\approx$112\,000. We expect that about 500 of them
may be discovered during a 4 yr long LISA mission.

In addition, we found that within 1 Kpc around the Sun, there might be
approximately 130 He-star AM~CVns with \porb$\aplt$(42-43)\,min. (assuming that they follow space distribution (\ref{eq:space})). This is the lower limit of the number of He-star AM~CVn systems, since we were not
able to trace evolution for \porb$\apgt$43\,min. In fact, we lost the majority of the systems, since evolution
slows down.
One hundred and thirty stars correspond to the space density of $3.1\times10^{-8}\, {\rm pc}^{-3}$. This number may be compared to a 2$\sigma$ limit on the space density
of \ams\ based on {\it Gaia} DR2 data $\rho > 7\times10^{-8}$\,\pc3 \citep{2018A&A...620A.141R}.
 Keeping uncertain selection effects in mind, serendipitous discoveries of \am\ stars,  taking into account, on the one hand,  that He-star \am\ stars may evolve much longer than we can account for, but, on the other hand, WDs might explode without leaving bound remnants or disrupting the binary, we deem that this result  does not contradict the finding of only three candidate  He-donor  \ams\ in the sample of known objects.

Published estimates  for the population of \ams\ were obtained under  different assumptions.
The most important computed parameters
are distributions of the initial binaries over the IMF of the primaries, mass ratios of components, orbital separations and eccentricities, the spatial distribution of binaries, star formation history (SFH),  the CE formalism, the efficiency and consequences of accretion, and the treatment of the evolution of a He star.
In addition, the estimates of the number of \am\ stars that might  be detected by  LISA vary as the project itself and  suggested data processing evolve.
We list below the results of some computations with available data on assumed detection details.

\citet{1996MNRAS.280.1035T} found  the birthrate of He-donor \am\ stars $\nu_{\rm He-AM}=4.9\times10^{-3}$\,\pyr\ for CE
efficiency \ace=1 (when considering common envelopes, they set the donor envelope binding energy factor $\lambda=1$). The birthrate  declined to 0 for \ace=0.1. The reason is clear: future He-donors are formed via CE. Roughly, post- and pre-CE separations are related as
$a_f \propto \ace \times a_i$ (see below). For small \ace, all possible precursors merge in CE. If \ace\ exceeds some limit, all
post-CE
systems are so wide that He in the cores of potential donors
is exhausted  before RLOF. The evolution of He-donor \am\ stars was integrated, assuming constant
$\dot{M}=3\times10^{-8}$\,\myr.
For \ace=1, the number of
objects was estimated as $N_{\rm He-AM}=4.9\times10^5$ to $1.9\times10^5$.
The reason for the rather small
number of objects was the assumption that, depending on \mwd\ and \mdot,  the accreted layer may either detonate after the accumulation of 0.2\,\ms\ of He and destroy the WD in a supernova (SN)  or it may strongly
expand and an R~CrB star may form.

\citet{nele01b} and \citet{nele04}, who used  code SEBA, varied SFR(t), several assumptions on stellar evolution, the Galactic model, and age.  As opposed to most other studies, the CE formalism based on the conservation of angular momentum was used for stars with non-compact components. Most important, it was assumed that an accreted He layer may detonate after the accretion of either 0.15\,\ms\ or 0.30\,\ms. Within the range of different assumptions, $\nu_{\rm He-AM}$ varied
from $0.27\times10^{-3}$ to $1.6\times10^{-3}$,  $N_{\rm He-AM}$ varied from
$1.1\times10^7$ to $3.1\times10^7$, and surface density $\rho$ exceeded $2\times10^{-5}$\,\pc3.

\citet{2012ApJ...758..131N}, using the updated version of SEBA \citep{2011arXiv1101.2787T} and varying
assumptions on the occurrence of double detonations and the Galactic model, estimated
$N_{\rm He-AM}$ as 0 to $1.12\times10^7$. The number of detectable He-star \am\ systems in the most optimistic case (5\,Mkm detector) was only about 80.
\begin{figure}[t]
\includegraphics[trim={-1.6cm 0 0 0}, width=0.7\columnwidth,angle=-90]{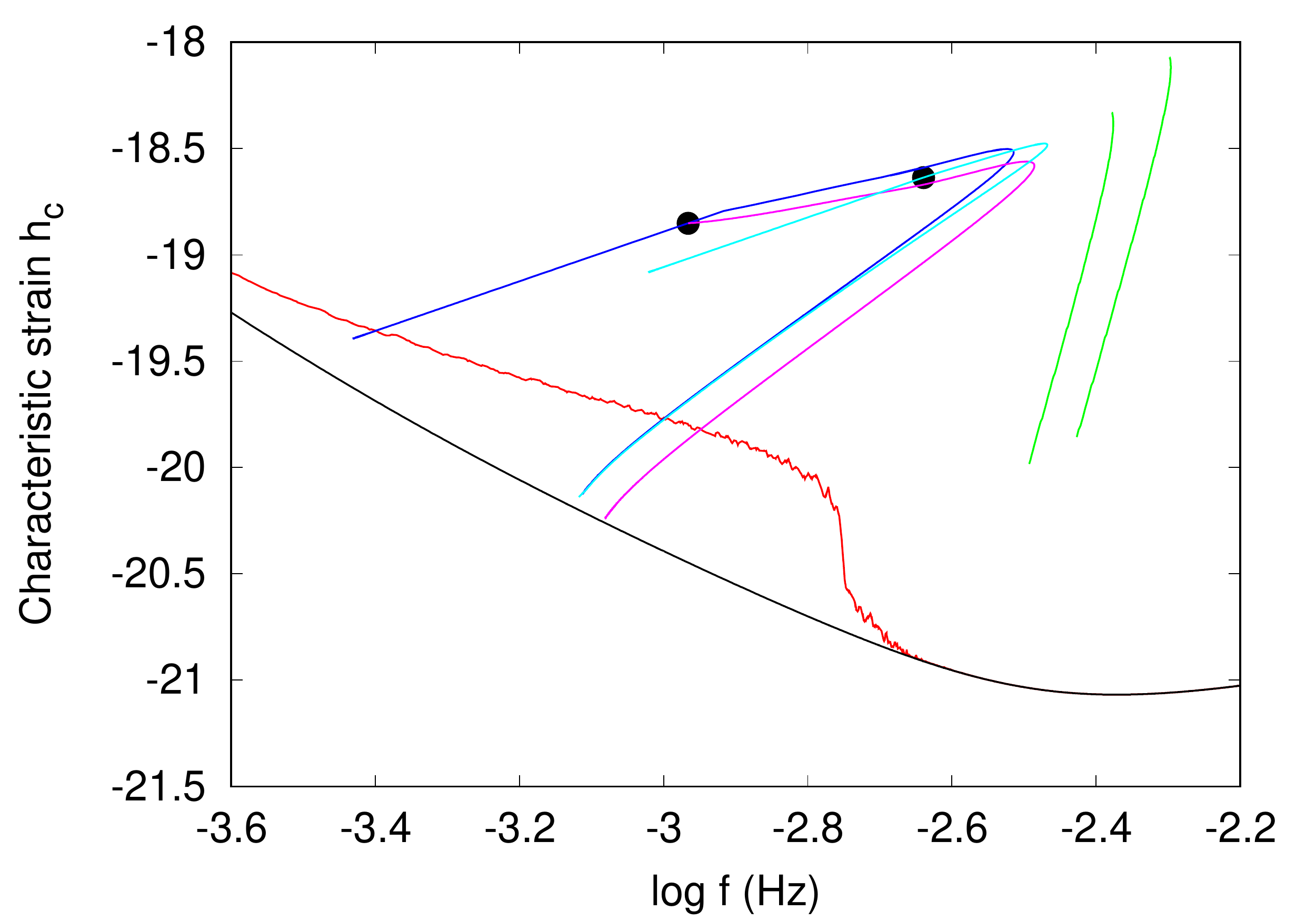}
\caption{Comparison of the tracks computed in the present study and the tracks computed using $M-R$ relations, as in \citet{nele01b}.
The blue line represents the track for
($M_{\rm He,0}+M_{\rm WD,0}, P_0$) = (0.43+0.87)\,\ms, 90\,min.; the
magenta line is the part of the track for the same system under the assumption of non-conservative mass exchange; and the
cyan line represents the track for
($M_{\rm He,0}+M_{\rm WD,0}, P_0$) = (0.33+0.72)\,\ms, 35\,min.
Black dots mark RLOF.
Green lines indicate the tracks computed using analytic $M-R$ relations
for the systems
($M_{\rm He,0}+M_{\rm WD,0}, P_0$) = (0.33+0.72)\,\ms, 35\,min. and
($M_{\rm He,0}+M_{\rm WD,0}, P_0$) = (0.43+0.87)\,\ms, 90\,min.
(left to right, respectively).
 The black line is the LISA noise line, and the
 red line indicates the gravitational waves' foreground+LISA noise.}
\label{fig:tracks}
\end{figure}

The basic difference between the present study and the studies involving  SEBA
is the treatment of the evolution of He stars.
In SEBA, an analytic approximation to the $M-R$ relation was used,
minimum  masses  of the donors were 0.007\,\ms, and the Galactic age was set to 13.5\,Gyr. The
$M-R$ relation roughly approximated the post-period minimum part of the donor track for the system (0.5+1.0)\,\ms\ in which the donor overfilled the Roche lobe almost unevolved \citep{1989SvA....33..606T}.
For the sake of comparison, we computed the evolution of the
($M_{\rm He,0}+M_{\rm WD,0}, P_{\rm 0}$)=(0.43\ms+0.87\ms, 90 min.) and
($M_{\rm He,0}+M_{\rm WD,0}, P_{\rm 0}$)=(0.33\ms+0.72\ms, 35 min.) systems using
this $M-R$ relation.
Computations   were continued until $M_{\rm He}\approx0.007$\,\ms, as in
\citet{nele01b}; readers can refer to  Fig.~\ref{fig:tracks} for more information.
This Figure clearly shows the major difference to the tracks computed by a full evolutionary code: a shift to larger frequencies and strain. The lifetime of the systems in the  region of the $f-h_c$ diagram
above the LISA sensitivity line
is 41.4\,Myr and 67.5\,Myr,
by about 30\% and 50\% longer, respectively, than that of the systems computed by an evolutionary code.
These factors, along to
the higher birthrate and  different Galactic model,  may be partially responsible for the higher number of potentially
detectable systems  and their higher frequencies in the models of
Nelemans and his coauthors.

Other studies of \am\ stars and their detection neglected He-star systems, as they have been deemed unimportant sources for LISA compared to detached binary WDs and DD \am\ stars. We only list below some estimates of detection rates.
\citet{nele04}  estimated the total number of detectable DD  \ams\ as
$\approx$11\,000 for a 5 yr mission and S/N>5;
\citet{2010ApJ...717.1006R}, assuming a 1 yr long mission, S/N>5 and a  5 Gm arm length for the detector
found  N=5300 sources;
\citet{2012ApJ...758..131N} estimated the number as N$\aplt$2000;
\citet{2013MNRAS.429.1602Y} found 8010, 19820, and 3840 objects for quasi-exponential, constant, and instantaneous SFH, respectively, after a 1 yr of
integration and S/N>3;
\citet{krem17a} found 2700 sources, requiring  S/N>5 and
negative chirp $<0.1\,{\rm yr^{-2}}$;
\citet{2018ApJ...854L...1B} provide N$\sim$3000 as an average of different assumptions on common envelope parameters for a 4 yr long  mission. Keeping all of the uncertainties in BPS in mind,
as well as
different detection criteria, all estimates are, in fact, in the same range.

\citet{2020ApJ...904...56G}, using  a simplified BPS and a grid of tracks
\citep{2019A&A...629A.134G}, estimated the number of stripped (0.3 -- 2.5\,\ms) He stars with WD
companions in the Galaxy as $\sim90\,000$ and suggested  that 15\% of these systems are currently in the mass-transfer stage.
  \citet{2020ApJ...904...56G} do not present
$M_{\rm WD}-M_{\rm d}$ relation for the binaries that started mass transfer, but
a naked-eye comparison of Fig.~\ref{fig:mwd_mhe_p} from the present paper and Figs.~2 and 3 from
\citet{2020ApJ...904...56G} suggests that no more than about 30\% of binaries
($\approx$4000) from the
`interacting' sample  will experience
stable  mass transfer
($M_{\rm d} \aplt 1$\,\ms, $P_{\rm orb} \aplt $ 1\,hr). This
number is still commensurate with our estimate of $\simeq$14\,800 pre-\am\ stars, keeping
differences as to assumptions about the
initial distribution of binaries over masses, periods, CE parameters, and evolutionary codes in mind.
\citet{2020ApJ...904...56G} estimate that under extremely  favourable assumptions,
LISA will be able to detect, within a 10 yr mission,  about 100 He-star+WD systems with S/N>5.
However, realistic  assumptions suggest  numbers $\aplt$10.

\subsection{Dependence on assumptions}
While the existence of DD  \am\ stars
is  beyond a doubt, there are some questions concerning their formation and fate.
This is associated, foremost,
with the orders of magnitude  discrepancy of  observed and predicted  space densities of the objects $\rho$. Observations suggest
$\rho{>} 7 \times 10^{-8}$\,\pc3 \citep{2018A&A...620A.141R}, which is at least an order of magnitude
less than the numbers listed above.
Among the major unsolved problems is the stability of mass exchange immediately after RLOF by He WDs since, in the case of inefficient spin-orbit coupling, most systems should merge
\citep{nele01b,2004MNRAS.350..113M,2016ApJ...824...46B}.
Existing observational data on  the slow rotation of two AM~CVn stars
\citep{2016MNRAS.457.1828K} are too scarce for any conclusions to be drawn.

\citet{2015ApJ...805L...6S} noted a problem common to other CVs as well \citep[see also][]{2021ApJ...923..100M,2022arXiv220506283S}.
In the initial stages of RLOF, H-rich matter is transferred,
leading to the classical novae outbursts. In the \am\ stars, a later  transfer of He may cause
outbursts. If these events result in the formation of envelopes that engulf both components,
dynamical friction may shrink the orbits, enhance the mass-transfer rate, and finally lead to the merger of components.
Just-formed `stripped stars' possess H-rich envelopes \citep{1970AcA....20..213Z}. Thus, they may first experience a series of H novae, if $\dot{M}$ is appropriate and, later, outbursts of He burning, also accompanied by the formation of common envelopes. However, these inferences should be confirmed by modelling WD motion inside postulated  envelopes.

We assumed that mass exchange in pre-\am\ and \am\ systems is conservative. This is
a certain simplification, since it is known that He accretion onto WDs at the rates close to the range of expected accretion rates in \am\ stars may result in thermonuclear outbursts of a different
strength in the layer of accreted He
\citep[e.g.][]{ 1980ApJ...242..749T,1982ApJ...253..798N,
1982ApJ...257..780N,1991ApJ...370..615I,
2011ApJ...734...38W,2014MNRAS.445.3239P},
ranging from weak flashes to detonations, potentially initiating sub-Chandrasekhar  SNe\,Ia
via a mechanism of double-detonation \citep{livne90} or,  for example, faint peculiar SN\,Iax due to deflagration  in a He layer \citep{2009A&A...493.1081J}. If SN disrupts the binary, a single helium-rich object may be formed. \citet{2015Sci...347.1126G} suggested  that runaway helium star US~708 is a remnant of a binary disrupted by a double-detonation sub-Chandrasekhar SN. Two further candidates of the same class were recently suggested by
\citet{2022A&A...663A..91N}. We note, however, that an attempt to model a population of high- and hypervelocity He stars in the
latter paper hints to a very low rate of events that disrupt He-star+WD binaries.

In the  strong flashes, accretor loses accumulated  He layer partially or completely.
However, the issue of possible detonations and matter retention efficiency
in flashes has not  been solved yet, especially since the character of thermonuclear flashes  depends
on the rotation of the accretor.  In the sample of pre-\am\ stars generated by BPS, we  found no system satisfying all conditions for accumulation of the He layer prone to detonation of rotating WDs formulated  by \citet{2019A&A...627A..14N}.
However, as an illustration of the possible influence of the loss of matter, we present in Fig.~\ref{fig:tracks}
the track for  the
($M_{\rm He,0}+M_{\rm WD,0}, P_{\rm 0}$)=(0.43\ms+0.87\ms, 90 min.)
system computed under an extreme assumption that all accreted  matter is lost by WD by isotropic re-emission.
The experiment shows two effects. First, a decrease in the total mass of the system results in reduced strain (see Eqs.~(\ref{eq:hc}) and (\ref{eq:chirp})), but the difference is not significant.
Second, since isotropic re-emission slows down widening of the system, the non-conservative binary spends about 60\,Myr  above confusion limit, which is by 50\% longer than the time spent by its
 conservative counterpart.
The combined effect would be an increase in the number of AM CVn stars above the confusion limit, but with slightly weaker signals.

To compare this with
the ($M_{\rm He,0}+M_{\rm WD,0}, P_{\rm 0}$)=(0.43\ms+0.87\ms, 90 min.) system discussed above, we plotted in Fig.~\ref{fig:tracks} the track for a more typical system
($M_{\rm He,0}+M_{\rm WD,0}, P_0$) = (0.33\ms+0.72\ms, 35\,min.), also assuming the distance of 1\,Kpc. The system is observable as a detached He-star+WD binary  for 8.9\,Myr. It spends about 89\,Myr above foreground, that is 3 times longer than the more massive system.
However, at a given frequency, the signals are quite comparable.
At the orbital frequency $f\approx 10^{-3}$\,Hz, characteristic strain  $h_c$ becomes lower than the foreground level. The mass of the donor at this instant declines to $\approx$0.045\,\ms, and the mass exchange rate becomes $1.7\times10^{-9}$\,\myr, that is to say the star should still be relatively bright. Further evolution of the system as an \am\ star, which we were able to trace with STARS,  lasted for $\Delta t \approx$
334\,Myr. The mass of the last donor model is $\approx$0.026\msun, and the orbital period of the system is 43.7\,min.

The greatest uncertainty in the
BPS is the  treatment of common envelopes
\citep[see][for the latest review]{2020cee..book.....I}. It
is still an unsolved 3D problem and, therefore,  simple energy or angular momentum
balance considerations were applied.
 Since the post-CE separation of components
$a_f$ was evaluated using the energy balance,
\begin{equation}
\frac{a_f}{a_i}=\frac{M_c M_2}{M_d}\,\frac{1}{M_2+2M_e/(\alpha_{ce} \lambda r_L)},
\label{eq:ce}
\end{equation}
where $M_d$ is mass of the donor, $M_2$ is the mass of the accretor, $M_c$ is the mass of the donor core, $M_e$ is the mass of the donor envelope,
$\ace$ is the so-called common envelope efficiency, $\lambda$ is the binding energy parameter of the donor envelope, and
$r_L$ is the Roche lobe radius. The most problematic  term in  Eq.~(\ref{eq:ce}) is \al. It is evident that $\ace\leq 1$,  unless highly uncertain additional energy sources \citep[see][]{2020cee..book.....I} are invoked and \ace\ should be `individual'  for all binaries. In its own turn,
$\lambda$ depends on the evolutionary stage of the
star, core-boundary definition, possible account of terms, other than gravitational binding energy.  Since  $a_f/a_i \propto \al$, the attempts to evaluate $\ace$ from the empirical data, in fact, provide this product, but not \ace, unless{ some assumptions} as to $\lambda$ were made.
Theoretically, the run of $\lambda$ along evolutionary tracks may be evaluated for certain sets of assumptions. As shown, for example, by
\citet{2000A&A...360.1043D} and \citet{2011ApJ...743...49L}, for intermediate-mass stars, precursors of components in He-star AM~CVns,
$\lambda$ is about 0.2-0.4 in the RGB stage and it becomes closer to 1 in the AGB stage. Keeping in mind that the formation
of He-star AM CVns invokes two common envelope episodes (Fig.~\ref{fig:evol}) and  uncertainties in the derivation of \ace\ and $\lambda$, we consider \ace=1   as a reasonable and  conservative assumption.

As a test, we performed three runs of BSE for $10^6$ initial systems, assuming \ace=0.5, 1, and 4. The obtained numbers of precursors of He-star AM~CVns
were 275, 5296, and 21670. This is understandable: by virtue of the relation $a_f/a_i \propto \al $, more systems merge in common envelopes for small
\ace\ and many more survive for, probably nonphysical, high \ace. We note that even for \ace=0.5, the
number of those potentially observed by LISA stars would be small, but not negligible
(in our computations for $10^5$ initial binaries, there were 524 progenitors for \ace=1).

\subsection{Observed sdB+WD binaries}
\label{sec:obs_sd}
Along to the He-donor \am\ stars, we estimated the number of their immediate precursors -- detached stripped He-star+WD binaries.
It is clear that the lifetime in the pre-\am\ stage is short, and the number of precursors is small, close to 14\,800. About 80 of them may be detected by LISA and they should be among
the strongest sources (Fig.~\ref{fig:f_hc}). In observations, they should be identified with detached sdB+WD systems.

We plotted in Fig.~\ref{fig:f_hc} positions of well-studied sdB+WD binaries with estimated distances. It is worth considering their
destiny and possible relation to the \am\ stars.

Detached binary CD-30$^{\circ}$11223
\citep{2012ApJ...759L..25V,2013A&A...554A..54G} has
$M_{\rm sdB}\approx$0.51\,\ms, $M_{\rm WD}\approx$0.75\,\ms, and $P_{\rm orb}$=70.5\,min.;
PTF~J2238+7430 \citep{2022ApJ...925L..12K} has $M_{\rm sdB} \approx 0.383$\,\ms,
$M_{\rm WD} \approx 0.725$\,\ms, and $P_{\rm orb}$ = 76.3 min.; and OW~J08153241
\citep{2022MNRAS.513.2215R} has $M_{\rm sdB}\approx$ 0.343\,\ms, $M_{\rm
WD}\approx$0.707\,\ms, and $P_{\rm orb}\approx$ 73.7\,min.
\citet{2019A&A...627A..14N} have shown that in the systems with combinations of
donor and rotating WD masses, as in these systems, detonation of an accreted
He layer does not occur. Rather, deflagrations and ejection of some matter may
be expected. Thus these binaries may become \am\ stars. For PTF~J2238+7430, our conclusion
is in disagreement with that of \citet{2022ApJ...925L..12K}, who expect double-detonation
after the accumulation of 0.17\,\ms\ of He, but they did not take rotation
effects into account.

HD265435 is a massive ($M_{\rm sdB} \approx 0.62$\,\ms, $M_{\rm
WD} \approx 0.91$\,\ms), wide (\porb$\approx$90.1\,min.) system
\citep{2021NatAs...5.1052P}. Large \porb\ suggests that the RLOF by sdB will
happen when He in its core will be considerably exhausted. Because of the high mass
of the donor, the presumably low abundance of He in the core, and the expected continuation of
He burning after RLOF, we expect that HD265435 will not become an AM CVn star,
but its donor will  merge with the companion, possibly,
with a \sna. A similar conclusion was also reached  by
\citet{2021NatAs...5.1052P}.

As described in \S 3, post-RLOF evolution of He stars depends on their mass and
degree of exhaustion of He in their cores. An additional factor, which defines
the possible outcome of accretion of He onto WDs, is rotation.

The semi-detached system ZTF J2055+4651 \citep{2020ApJ...891...45K}, with $M_{\rm sdB}
\approx 0.4$\,\ms\ and $M_{\rm WD} \approx 0.68$\,\ms, has large $P_{\rm orb}
=56.35$ min. The presence of H in the spectrum means that the subdwarf
overflowed the Roche lobe close to the current $P_{\rm orb}$. This implies that He
in the core of the subdwarf is almost exhausted and it will evolve into a hybrid WD
and merge with the companion, as also suggested by \citet{2020ApJ...891...45K}.
The system ZTF~J2130+4420 \citep{2020ApJ...898L..25K} is similar to
ZTF~J2055+4651: $M_{\rm sdB}\approx 0.337$\,\ms, $M_{\rm WD}\approx 0.545$\,\ms, and
$P_{\rm orb}$=39.34\, min. There is H in the spectrum too and it may be expected
that its evolution must be similar to that of ZTF~J2055+4651.

The most recently discovered system J1920-2001 \citep{2022MNRAS.515.3370L}, with
$M_{\rm sdB}\approx 0.337$\,\ms\ and
$M_{\rm WD}\approx 0.545$\,\ms,
differs from the above-mentioned systems by a large
\porb=3.4946\,hr. The position of the subdwarf in the $T_{eff}-\log\,g$ diagram suggests that it is in the He-shell burning stage. After completion of this stage, it will turn into a WD and merge with its companion within $\simeq 1$\,Gyr, as estimated by  \citet{2022MNRAS.515.3370L}.

\subsection{The number of stripped He stars in the Solar vicinity}
At the suggestion of the referee, we compared the model number of stripped He stars in the 1-Kpc vicinity of the Sun with the number of objects in the same region in the catalogue of known hot subdwarfs \citep{2020A&A...635A.193G}. In fact, the solution to such a problem requires {full} population synthesis of subdwarfs and is far beyond the aim of the present paper.

The stars in Geier's catalogue have parallaxes $\pi$\  from {\sl GAIA} DR2. To obtain the distances to the stars $d$,  we cross-correlated this catalogue with that of \citet{2018AJ....156...58B},  which provides {\sl Gaia} distances corrected
for non-linearity of $\pi \to d$ transformation.

Assuming Galactic SFR after \citet{2011MNRAS.417.1392Y} and using the same assumptions as for modelling  an \am\ population, we estimated,
using BSE, the birthrate $\nu$\ and current Galactic number
$N$ of detached  He-star+WD systems
 ($\nu\approx5.8\times10^{-4}$\,\pyr\ and
$N\approx40$),
He-star+MS star binaries
($\nu\approx1.9\times10^{-3}$\,\pyr\ and $N\approx130$
for $M_{\rm He}\le1.5$\,\ms),
and single sub-dwarfs that formed by the merger of He WDs
($\nu\approx9.4\times10^{-5}$\,\pyr, $N\approx35$).
The number of obtained model objects ($\approx200$) is 5 times lower
than the number of objects within 1 Kpc from the Sun in the
subdwarfs catalogue.

However, observed hot subdwarfs are a
 mixture of objects (see, e.g. comprehensive review by \citet{2016PASP..128h2001H}), possibly forming also via channels different from those listed above. We note that  we did not consider a possible merger of red giants and low-mass main-sequence stars in the common envelopes.
As suggested by \citet{2008ApJ...687L..99P}, such mergers may result
in the formation of single hot subdwarfs. The solution to the problem requires
3D modelling, which is still beyond our current capabilities.
In our model, such mergers occur at the rate
$\nu=1.8\times10^{-2}$\pyr. If this channel really produces predominantly hot subdwarfs and if their typical lifetime is  (200 -- 300)\,Myr, as that of the
subdwarfs that formed via `stripping' channel, this
scenario may occur to be the main route for formation of single hot
subdwarfs.
Keeping in mind that within 1~Kpc from the Sun reside about 0.1\% of all hot subdwarfs formed via the `stripping' channel, formation via merger  may also resolve the
problem of `deficiency' of model stars respective to the catalogued ones.

We also remind readers that there is still an unresolved hypothesis about single subdwarfs  precursors -- `hot flashers' \citep[see, e.g.][and references therein]{1996ApJ...466..359D} -- or suggestions that subdwarfs may be formed
due to an interaction of red giants with brown dwarfs
\citep{2010Ap&SS.329...25N} and even planets \citep{1998AJ....116.1308S}.

\subsection{Conclusions}
To summarise, we explored the formation of short and moderate period (\porb$\aplt$43~min.) \am\ stars with He donors, conjoining fast BPS for their progenitors with detailed evolutionary computations for the \am\ stage itself. We found that the number of such systems in the Galaxy -- if their components do not merge due to tidal friction in the novae envelopes in the early stages of mass transfer and if they are not destroyed by He detonations -- may amount to $\simeq$112\,000. About 500 of them may be detected by LISA with S/N>5 during a 4-yr mission. In addition, LISA may detect up to  80 of their immediate precursors.

Helium-star AM~CVns were modelled separately from DD in several papers using code SEBA \citep{nele01b,nele04} and its clones only. Since we did not  cover the entire range of \porb, it is impossible to compare predicted numbers directly.  Nevertheless, it is clear that we expect  a much lower number of \ams,\, mostly because of the difference between the results of
evolutionary computations for binaries and results, based on analytic $M-R$ relations.
A low rate of predicted detections of \ams\ in GWR in our study may be justified
since the main limiting factor is the confusion limit.

The scarcity of He-star \ams\ in the total sample of the observed \am\ stars
may mean that unrecgnized recognized selection effects still exist.
As well,  the problem of possible merger of components due to tidal friction in the envelopes ejected in strong flashes of nuclear burning of
accreted H and He still remains open.
The same concerns possible explosions of accreting WDs as SNe or the decay of
binaries due to mass ejection in the outbursts of He burning.
Possible non-detection of these stars by LISA may confirm these assumptions. The deficiency of He-donor AM CVn stars may also point to a low value of the product of `common envelope efficiency' and binding energy parameter $\al$. A much lower number of model hot subdwarfs within 1~Kpc from the Sun compared to the number of them with {\sl Gaia} distances, catalogued by \citet{2020A&A...635A.193G}, may indicate that there are other scenarios of formation of single subdwarfs, apart from the merger of WDs or the disruption of binaries. This problem definitely deserves a dedicated study.

\vskip 0.5cm
{\small We are grateful to the referee for her/his insightful comments and helpful suggestions. The authors acknowledge fruitful discussions with V. Korol, D. Kovaleva, G. Nelemans, L. Piersanti, K. Postnov, A. Rastorguyev. We are grateful to P.P. Eggleton for providing evolutionary code and for comments on its usage. S. Toonen is acknowledged for providing data on observations-driven confusion limit. H.-L. Chen is acknowledged for mindful discussion of the BPS with W.-M. Liu. W.-C. Chen is acknowledged for  the inspiration of this study. This study was supported by the National Natural Science Foundation of China (under grant Nos. U2031116 and 11733009), and the Research Start-up Funding Project of Shangqiu Normal University. A. Kuranov  was supported by the Interdisciplinary science and education school of M.V. Lomonosov Moscow University "Fundamental and applied space studies".  This research has made use of NASA's Astrophysics Data System.}

\bibliographystyle{aa}
\bibliography{paper}
\end{document}